\documentclass[aps,prb,twocolumn,groupedaddress,superscriptaddress,longbibliography]{revtex4-2}

\usepackage{amsmath}
\usepackage{amssymb}
\usepackage{graphicx}
\usepackage{dcolumn}
\usepackage{cases}
\usepackage{epstopdf}
\usepackage[hidelinks,colorlinks=true,linkcolor=blue,citecolor=blue,anchorcolor=black,filecolor=cyan,menucolor=blue,runcolor=cyan,urlcolor=blue]{hyperref}
\usepackage{bm}
\usepackage{subfigure}
\usepackage{makecell}
\usepackage{color}
\usepackage{txfonts}
\usepackage{bookmark}
\allowdisplaybreaks[4]
\usepackage{lineno}

\makeatother

\begin{document}

\title{Exact surface energy of the Hubbard model with nonparallel boundary magnetic fields}

\author{Pei Sun}
\affiliation{School of Physics, Northwest University, Xi'an 710127, China}
\affiliation{Peng Huanwu Center for Fundamental Theory, Xi'an 710127, China}
\author{Yi Qiao}
\affiliation{Peng Huanwu Center for Fundamental Theory, Xi'an 710127, China}
\affiliation{Institute of Modern Physics, Northwest University, Xi'an 710127, China}
\author{Tao Yang}
\email{yangt@nwu.edu.cn}
\affiliation{Peng Huanwu Center for Fundamental Theory,  Xi'an 710127, China}
\affiliation{Institute of Modern Physics, Northwest University, Xi'an 710127, China}
\affiliation{Shaanxi Key Laboratory for Theoretical Physics Frontiers, Xi'an 710127, China}
\author{Junpeng Cao}
\email{junpengcao@iphy.ac.cn}
\affiliation{Peng Huanwu Center for Fundamental Theory,  Xi'an 710127, China}
\affiliation{Beijing National Laboratory for Condensed Matter Physics, Institute of Physics, Chinese Academy of Sciences, Beijing 100190, China}
\affiliation{School of Physical Science, University of Chinese Academy of Sciences, Beijing 100049, China}
\affiliation{Songshan Lake Materials Laboratory, Dongguan, Guangdong 523808, China}
\author{Wen-Li Yang}
\affiliation{School of Physics, Northwest University, Xi'an 710127, China}
\affiliation{Peng Huanwu Center for Fundamental Theory,  Xi'an 710127, China}
\affiliation{Institute of Modern Physics, Northwest University, Xi'an 710127, China}
\affiliation{Shaanxi Key Laboratory for Theoretical Physics Frontiers, Xi'an 710127, China}

\date{\today}

\begin{abstract}
{
In this study, we explore the precise physical quantities in the thermodynamic limit of the one-dimensional Hubbard model with nonparallel boundary magnetic fields based on the off-diagonal Bethe ansatz solution. A particular emphasis is placed on the half-filling condition to investigate the distinct patterns of Bethe roots in the reduced Bethe ansatz equations for different boundary parameters. The ground state of the system can be divided into five regions according to the distribution of Bethe roots. By analyzing these patterns, we calculate the densities of states, ground state energy density, and surface energy. The results reveal the existence of stable boundary bound states, which are dependent on specific constraints regarding the boundary magnetic fields.}
\end{abstract}
\maketitle

\section{Introduction}\label{sec1}

The Hubbard model plays a crucial role in describing strongly correlated electronic systems, making it one of the most significant models in this field. This model facilitates the study of various fascinating phenomena, including the Mott transition\cite{bib01}, high $T_c$ superconductivity\cite{bib02}, quantum spin liquids\cite{bib03}, and quantum entanglement\cite{bib04,bib05,bib-add}. In addition, the research on this model has led to the development of several noteworthy methods such as the Bethe ansatz\cite{bib11}, Monte Carlo simulations\cite{bib-taoyang-add01} and tensor network \cite{bib-taoyang-add02}. In one dimension, the Hubbard model can be exactly solved\cite{bib07}, which has led to significant advancements in our understanding of the model. The excitation spectrum\cite{bib08} of the one-dimensional Hubbard model and the R-matrix\cite{bib06, bib09} constructed as a foundational concept for the algebraic Bethe ansatz have provided a basis for the development of nested schemes\cite{bib10,bib11,bib12,bib13} that enable a deeper exploration of the model.

The integrable open boundary Hubbard model is a significant topic of study, alongside the periodic boundary condition. This model can be achieved by introducing two boundary magnetic fields to the system. When these boundary magnetic fields are parallel, the system exhibits $U(1)$ symmetry, resulting in diagonal boundary reflection matrices\cite{bib14}. The diagonal integrable boundary reflection matrices have been obtained through the solution of the reflection equation\cite{bib15,bib16,bib17,bib18,bib19,yang-25}. Several research papers have focused on the boundary effects induced by the boundary fields $n_{1\uparrow}+n_{1\downarrow}$ and $n_{L\uparrow}+n_{L\downarrow}$\cite{bibsur01, bibsur02}. Additionally, a study on the boundary scattering of the open Hubbard chain has been conducted\cite{bibsur03}. Various intriguing aspects, such as the ground state energy, elementary excitations, and magnetization, have also been examined\cite{bibsur04, bibsur05, bib23}. A recent study showed that in addition to the ground-state phase transition, boundary fields can induce a boundary eigenstate phase transition in the spin-1/2 Heisenberg chain\cite{natan,PRB.107.224412}.

In the case where the Hubbard model exhibits integrable non-diagonal boundary reflection matrices, indicating the presence of two unparallel boundary magnetic fields, the $U(1)$ symmetry of the system is violated, resulting in a lack of conservation for the number of electrons with fixed spin after the boundary reflection. However, the conventional Bethe ansatz method is not applicable in this scenario as it becomes challenging to construct an appropriate reference state, which is a crucial step in the solving process. Consequently, the exact solution for this problem has remained a longstanding challenge in the field. Recently, a new method called off-diagonal Bethe ansatz has been proposed\cite{cao1}, resulting in the successful determination of the exact solution for the Hubbard model with non-diagonal boundary reflections\cite{bib20, bib21,bib22}. Nevertheless, calculating physical quantities in the thermodynamic limit remains a difficult task because of the inhomogeneous nature of the corresponding Bethe ansatz equations (BAEs) and the inability to apply the thermodynamic Bethe ansatz method.

In this paper, we investigate the thermodynamic limit of the Hubbard model with nonparallel boundary magnetic fields, with particular focus on the half-filling case. Our main objective is to analyze the impact of the inhomogeneous term of the BAEs on the ground state in the thermodynamic limit, as revealed through finite size scaling analysis. We divide the system into five regions based on the values of the boundary parameters. In each region, we observe different distributions of the Bethe roots of the reduced BAEs. Specifically, we determine the patterns of Bethe roots and study their densities, ground state energy density, and surface energies as the system size approaches infinity. Our findings indicate the presence of stable boundary bound states under certain constraints on the boundary magnetic fields. Additionally, we employ the density matrix renormalization group (DMRG) methods\cite{bib24,bib25,bib26} to verify the accuracy of our analytical results. We find strong agreement between the analytical and numerical results, confirming the reliability of our findings.

The paper is structured as follows: In Sec. \ref{sec2}, we present the
off-diagonal Bethe ansatz solutions for the model. Section \ref{sec4}
outlines the distribution of the Bethe roots in the ground state.
The finite-size scaling behavior of the ground-state energy
is investigated in Sec. \ref{sec3}. Section \ref{sec5} focuses on calculating
the ground-state energy and the surface energy resulting from
the boundary fields in the thermodynamic limit. Finally, in
Sec. \ref{sec6}, we summarize the obtained results and provide further discussions.

\section{Bethe ansatz solutions}\label{sec2}

We investigate the Hubbard chain with general boundary fields
\begin{eqnarray}\label{Ham}
  H&=&-t\sum_{\alpha=\uparrow, \downarrow } \sum_{j=1}^{N-1}[c_{j,\alpha}^\dag c_{j+1,\alpha}+c_{j+1,\alpha}^\dag c_{j,\alpha}]+U\sum_{j=1}^N n_{j\uparrow}n_{j\downarrow}\nonumber\\
  &+&
  h_1^-c_{1,\uparrow}^+c_{1,\downarrow}+h_1^+c_{1,\downarrow}^+c_{1,\uparrow}
  +h_1^z(n_{1,\uparrow}-n_{1,\downarrow})\nonumber\\
  &+&h_N^-c_{N,\uparrow}^+c_{N,\downarrow}+h_1^+c_{N,\downarrow}^+c_{N,\uparrow}
  +h_N^z(n_{N,\uparrow}-n_{N,\downarrow}),
\end{eqnarray}
where $c_{j,\alpha}^+$ and $c_{j,\alpha}$ are the creation and annihilation operators of electrons on $j_{th}$ site with
spin component $\alpha=\uparrow, \downarrow$, respectively. We have  $j=1, \cdots,i, N$, and $N$ is the total number of sites.
The hopping constant, the number of electrons and the on-site repulsive interaction among them are denoted by
$t$, $n_{j,\alpha}$ and $U$, respectively. The boundary magnetic fields are described by $\mathbf{h}_1=(h_1^x,h_1^y,h_1^z)$, $\mathbf{h}_N=(h_N^x,h_N^y,h_N^z)$ and $h_j^{\pm}=h_j^x\pm h_j^y$.

Combining the coordinate and off-diagonal Bethe ansatz methods, the eigen-energy $E$ of the Hamiltonian \eqref{Ham} can be expressed by\cite{bib21}
\begin{equation}\label{prime-energy}
  E=-2t\sum_{j=1}^{\bar{M}}\cos k_j,
\end{equation}
where the quasi-momenta ${k_j}$ satisfy the inhomogeneous BAEs
\begin{eqnarray}
	\label{BAE01-off}
  &&\frac{4(p-\sin k_j\varepsilon |\mathbf{h}_N|)}{1-\mathbf{h}_1^2e^{2ik_j}}
  \frac{(q+\sin k_j|\mathbf{h}_1|)}{e^{-2ik_j}-\mathbf{h}_N^2}=
  e^{-2ik_j N}\nonumber\\
  &&\quad\times\prod_{l=1}^M\frac{(\sin{k_j}+\lambda_l-\frac{\eta}{2})}
  {(\sin{k_j}+\lambda_l+\frac{\eta}{2})}
  \frac{(\sin{k_j}-\lambda_l-\frac{\eta}{2})}
  {(\sin{k_j}-\lambda_l+\frac{\eta}{2})},\\
  &&\quad j=1,\cdots, \bar{M}\nonumber,
\end{eqnarray}
\begin{eqnarray}  \label{BAE02-off}
  && \frac{\lambda_j+\frac{\eta}{2}}{\lambda_j-\frac{\eta}{2}}
  \frac{p+(\lambda_j-\frac{\eta}{2})\varepsilon |\mathbf{h}_N|}
  {p-(\lambda_j+\frac{\eta}{2})\varepsilon |\mathbf{h}_N|}
  \frac{q-(\lambda_j-\frac{\eta}{2})\varepsilon |\mathbf{h}_1|}
  {q+(\lambda_j+\frac{\eta}{2})\varepsilon |\mathbf{h}_1|}\nonumber\\
  &&\quad\times\prod_{l=1}^{\bar{M}}\frac{\lambda_j+\sin k_l+\frac{\eta}{2}}
  {\lambda_j+\sin k_l-\frac{\eta}{2}}
  \frac{\lambda_j-\sin k_l+\frac{\eta}{2}}
  {\lambda_j-\sin k_l-\frac{\eta}{2}}=\nonumber\\
  &&\quad
  \prod_{l=1}^M\frac{\lambda_j-\lambda_l+\eta}{\lambda_j-\lambda_l-\eta}
  \frac{\lambda_j+\lambda_l+\eta}{\lambda_j+\lambda_l-\eta}\\
  &&\quad-c\frac{\lambda_j+\frac{\eta}{2}}{p-(\lambda_j+\frac{\eta}{2})\varepsilon
  |\mathbf{h}_N|}
  \frac{2\lambda_j}{q+(\lambda_j+\frac{\eta}{2})\varepsilon
  |\mathbf{h}_1|}
  \nonumber\\
  &&\quad\times\prod_{l=1}^{\bar{M}}\frac{(\lambda_j+\sin{k_l}+\frac{\eta}{2})}
  {(\lambda_j-\lambda_l+\eta)}
  \frac{(\lambda_j-\sin{k_l}+\frac{\eta}{2})}
  {(\lambda_j+\lambda_l-\eta)},\nonumber\\
  && \quad j=1, \cdots, M,\nonumber
\end{eqnarray}
$M$ is the number of electrons with spin-down while $\bar{M}$ is the total number of electrons. We have $M=N_\downarrow$ and $\bar{M}=N_\uparrow+N_\downarrow$. In the derivation, we have used the following notations:
\begin{eqnarray}
&&\eta=-i\frac{U}{2t}, \quad p=i\frac{\mathbf{h}_N^2-t^2}{2t},\quad
 q=i\frac{t^2-\mathbf{h_1^2}}{2t}, \nonumber \\
&&\varepsilon=\frac{\mathbf{h}_1 \cdot \mathbf{h}_N}{|\mathbf{h}_1 \cdot \mathbf{h}_N|},\quad
 c=2(\varepsilon|\mathbf{h}_1||\mathbf{h}_N|-\mathbf{h}_1\cdot \mathbf{h}_N).
 \label{sun-constant}
\end{eqnarray}
For the sake of convenience, we set $t$ equal to 1 and $\varepsilon$ equal to 1 throughout the rest of this paper.

The BAEs \eqref{BAE02-off} are non-homogeneous, making it impossible for them to be solved by taking the logarithm and expressing the integration equation in the thermodynamic limit. Therefore, we will instead focus on the homogeneous counterpart of the BAEs \eqref{BAE01-off} and \eqref{BAE02-off}:
\begin{eqnarray}\label{BAE01}
  &&\frac{4(p-\sin k_j |\mathbf{h}_N|)}{1-\mathbf{h}_1^2e^{2ik_j}}
  \frac{(q+\sin k_j|\mathbf{h}_1|)}{e^{-2ik_j}-\mathbf{h}_N^2}=
  e^{-2ik_j N}\nonumber\\
  &&\times\prod_{l=1}^M\frac{(\sin{k_j}+\lambda_l-\frac{\eta}{2})}
  {(\sin{k_j}+\lambda_l+\frac{\eta}{2})}
  \frac{(\sin{k_j}-\lambda_l-\frac{\eta}{2})}
  {(\sin{k_j}-\lambda_l+\frac{\eta}{2})},\\
  &&\quad j=1,\cdots, \bar{M}, \nonumber
 \end{eqnarray}
 \begin{eqnarray}\label{BAE02}
  && \frac{\lambda_j+\frac{\eta}{2}}{\lambda_j-\frac{\eta}{2}}
  \frac{p+(\lambda_j-\frac{\eta}{2})|\mathbf{h}_N|}
  {p-(\lambda_j+\frac{\eta}{2}) |\mathbf{h}_N|}
  \frac{q-(\lambda_j-\frac{\eta}{2}) |\mathbf{h}_1|}
  {q+(\lambda_j+\frac{\eta}{2}) |\mathbf{h}_1|}\nonumber\\
  &&\times\prod_{l=1}^{\bar{M}}\frac{\lambda_j+\sin k_l+\frac{\eta}{2}}
  {\lambda_j+\sin k_l-\frac{\eta}{2}}
  \frac{\lambda_j-\sin k_l+\frac{\eta}{2}}
  {\lambda_j-\sin k_l-\frac{\eta}{2}}\\
  &&=
  \prod_{l=1}^M\frac{\lambda_j-\lambda_l+\eta}{\lambda_j-\lambda_l-\eta}
  \frac{\lambda_j+\lambda_l+\eta}{\lambda_j+\lambda_l-\eta},  \quad j=1, \cdots, M.\nonumber
\end{eqnarray}
Substituting the solutions of reduced BAEs (\ref{BAE01})-(\ref{BAE02}) into Eq.\eqref{prime-energy}, we obtain
\begin{equation}\label{half-eigenvalue}
  E_{hom}=-2\sum_{j=1}^{\bar{M}}\cos k_j.
\end{equation}

\begin{table*}
	\centering
	\caption{The patterns of Bethe roots at the ground state in different regions, where $N$ is the number of electrons and $M$ is the number of the Bethe roots, $\alpha=|\mathbf{h}_1|$ and $\beta=|\mathbf{h}_N|$. }
	\begin{ruledtabular}
		\begin{tabular}{cccc}
			$\text{Region}$ & $\{k_j| j=1,\cdots, N\}$ & $\{\lambda_j| j=1,\cdots, M=\frac{N}{2}\}$
			& $\alpha,\beta$\\
			\hline
			i & $k_j \in {\bf R} $ & $\lambda_j \in {\bf R} $ & $ 0<\alpha<h_0 \quad 0<\beta<h_0$\\
			\hline
			ii & $ k_j \in {\bf R}$ & $
			\makecell{\lambda_j \in {\bf R},  j=1,\cdots, \frac {N} {2}-1 \quad \lambda_{M}=i
				\left(\frac{1-\beta^2}{2\beta}-\frac{U}{4}\right)}$  & $ 0<\alpha<h_0 \quad h_0<\beta<1$\\
			\hline
			iii & $ k_j \in {\bf R}$ & $
			\makecell{\lambda_j \in {\bf R},  j=1,\cdots, \frac {N} {2}-1 \\
				\lambda_{M-1}=i
				\left(\frac{1-\alpha^2}{2\alpha}-\frac{U}{4}\right) \quad
				\lambda_{M}=i
				\left(\frac{1-\beta^2}{2\beta}-\frac{U}{4}\right)}$  & $ h_0<\alpha<1 \quad h_0<\beta<1$\\
			\hline
			iv & $\makecell{k_j \in {\bf R}, j=1,\cdots, N-1 \\
				k_N=\arcsin(i\frac{\beta^2-1}{2\beta})} $ & $\makecell{\lambda_j\in {\bf R} \quad j=1,\cdots, \frac{N}{2}-1 \\ \lambda_{M}=i
				\left(\frac{\beta^2-1}{2\beta}+\frac{U}{4}\right)}$  &
			$ 0<\alpha<1 \quad \beta>1 $ \\
			\hline
			v & $\makecell{k_j \in {\bf R}, j=1,\cdots, N-2 \\
				k_{N-1}=\arcsin(i\frac{\alpha^2-1}{2\alpha})
				\\
				k_N=\arcsin(i\frac{\beta^2-1}{2\beta})}$ & $\makecell{
				\lambda_j\in {\bf R} \quad j=1,\cdots, \frac{N}{2}-2 \\ \lambda_{M-1}=i
				\left(\frac{\alpha^2-1}{2\alpha}+\frac{U}{4}\right) \quad
				\lambda_{M}=i
				\left(\frac{\beta^2-1}{2\beta}+\frac{U}{4}\right) }$  & $ \alpha>\beta>1$ \\
		\end{tabular}
	\end{ruledtabular}
	\label{table01}
\end{table*}

Some remarks are in order. Firstly, $E$ represents the eigen-energy of the Hamiltonian \eqref{Ham}, with the condition that the boundary fields are non-parallel. If $c=0$, which can be achieved by ensuring that the two boundary magnetic fields satisfy certain constraints, then $E_{hom}$ is equal to $E$. However, for arbitrary values of the boundary parameters, $E_{hom}$ is not equal to $E$. The BAEs (\ref{BAE01}) and (\ref{BAE02}) represent the equations of the system with parallel boundary magnetic fields, and Eq.\,\eqref{half-eigenvalue} represents the corresponding energy. Secondly, by solving Eqs.\,\eqref{BAE01} and \eqref{BAE02}, we can obtain the values of $E_{hom}$. Alternatively, $E_{hom}$ can be obtained in another routine. The integrable boundary magnetic fields are parallel and are applied only in the spin sector. The strengths of these magnetic fields are represented by $|\mathbf{h}_1|$ and $|\mathbf{h}_N|$. Because of the $su(2)$ symmetry in the spin sector of the system, we can rotate the system in the spin sector and choose the direction of the magnetic fields as the $z$-direction. This observation provides an alternative numerical method for calculating the values of $E_{hom}$. The results obtained through the two methods are identical.

\section{Patterns of Bethe roots}
\label{sec4}
Now, we shall investigate the ground state, specifically at half-filling $N=\bar M$.
 The ground state energy are quantified by the BAEs. Thus we should solve the BAEs first.
The solutions of Bethe roots can be obtained by the singularity analysis of BAEs \cite{xiu-bib01, xiu-bib02, xiu-bib021}.
If a Bethe root is the complex number in the upper complex plane, substituting this Bethe root into BAEs, one will find that the left hand side of BAEs tend to infinity (or zero)
in the thermodynamic limit.
Then the right hand side of BAEs must also tend to infinity (or zero) to keep the BAEs hold.
Based on this principle, the Bethe roots form the strings.
We should note that if the length of string is zero, the corresponding Bethe root is the real number.
The Bethe roots include the bulk stings and discrete boundary strings.
The boundary strings correspond the table boundary bound states.
In the thermodynamic limit, the bulk strings give the density of roots
and the boundary strings characterized the corrections.

Because the boundary fields are applied on the spin sector, we only list the singularity analysis of BAEs (\ref{BAE02}).
The singular points of BAEs \eqref{BAE02} are
\begin{equation}\label{sun-h0}
\frac{1-|\mathbf{h}|^2}{2|\mathbf{h}|}-\frac{U}{4}, \quad \mathbf{h}=\mathbf{h}_1, \mathbf{h}_N.
\end{equation}

If $|\mathbf{h}|=1$, then the singular point degenerates to $-\frac{U}{4}$.
Substituting $|\mathbf{h}|=1$ into the BAEs (\ref{BAE02}), we find that the corresponding boundary parameters are not included in the BAEs and
the related terms equal to $\frac{\lambda_j+i\frac{U}{4}}{\lambda_j-i\frac{U}{4}}$, which is canceled by the structure factor $\frac{\lambda_j-i\frac{U}{4}}{\lambda_j+i\frac{U}{4}}$.
Neglecting the terms with boundary reflection and the boundary field on the other side of the chain, the BAEs (\ref{BAE02})) are the same as the ones for the periodic boundary condition.
Then we conclude that the critical point $|\mathbf{h}|_c=1$ characterizes the periodic structure of the scattering process of the quasi-particles.
In one periodicity of the quasi-particles moving,
the complete scattering and reflecting processes are the following. The particle $j$ moves from the left boundary to the
right boundary by scattering once with all the other particles. Then it is bounced back by the right boundary with a reversed momentum. The reflected particle moves
further from the right boundary to the left boundary by scattering once again with all the other particles. Finally it is bounced back by the left boundary and arrives at
its initial position. This process is described by the transfer matrix, which is generating function of the model Hamiltonian.

If  $|\mathbf{h}|=h_0$, the singular point degenerates to zero
\begin{equation}\label{sun-h0-02}
\frac{1-h_0^2}{2h_0}-\frac{U}{4}=0.
\end{equation}
Substituting $|\mathbf{h}|=h_0$ into the BAEs (\ref{BAE02}), we find that the boundary parameters are not included in the BAEs. However, the
structure factor is maintained, which can induce the boundary parameters independent string solution.
The resulting energy is the surface energy induced by the free open boundary (open, but without boundary fields).
This conclusion agrees with the fact that the surface energy of the system with boundary fields is the summation of the energy induced by the open boundaries and the one induced by the boundary parameters \cite{xiu-bib022}.

\begin{figure}%
	\centering
	\includegraphics[scale=0.65]{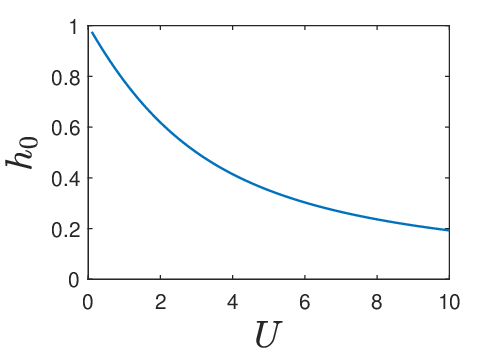}
	\caption{The value of critical field $h_0$ vs the on-site coupling $U$. $h_0$ satisfies $({h_0^2-1})/{2h_0}+{U}/{4}=0$ and $h_0>0$.}
	\label{fig000-h0}
\end{figure}

Based on the above analysis, we know that there are two critical values for the boundary fields. One is $|\mathbf{h}|_c=1$ and the other is $|\mathbf{h}|_c=h_0$.
The singular point $1$ characterizes the periodic structure of the scattering process of the quasi-particles,
and the singular point $h_0$ characterizes the free open boundary without external boundary magnetic field.
In these two cases, the boundary strings are the constants and do not depend on the boundary parameters.
The value of $h_0$ versus the on-site coupling $U$ is shown in Fig.\ref{fig000-h0}. From it, we see that $h_0$ is always smaller that 1.
According to this, we conclude that the boundary parameters can be divided into five regions: \begin{description}
\item[i] $0<|\mathbf{h}_1|<h_0$, $0<|\mathbf{h}_N|<h_0$;
\item[ii] $0<|\mathbf{h}_1|<h_0$, $h_0<|\mathbf{h}_N|<1$;
\item[iii] $h_0<|\mathbf{h}_1|<1$, $h_0<|\mathbf{h}_N|<1$;
\item[iv] $0<|\mathbf{h}_1|<1$, $|\mathbf{h}_N|>1$;
\item[v] $|\mathbf{h}_1|>1$, $|\mathbf{h}_N|>1$.
\end{description}

\begin{figure}%
	\centering
	\includegraphics[scale=0.65]{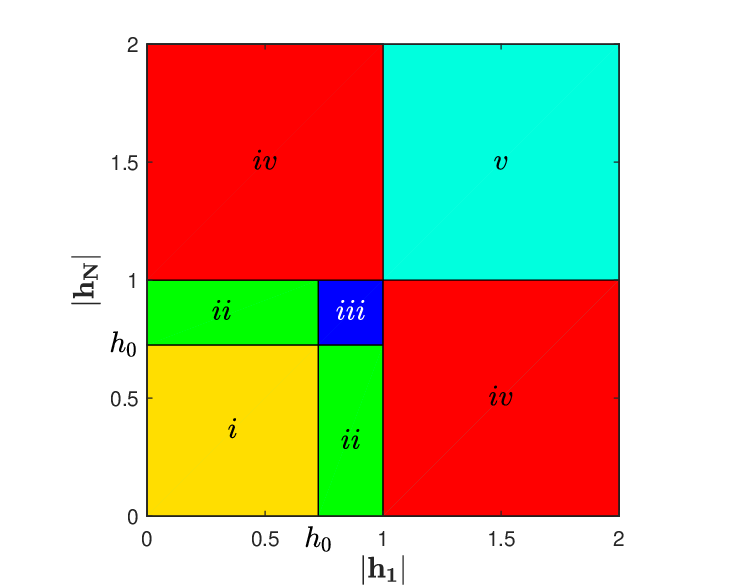}
	\caption{The different regions of the boundary parameters in the $\mathbf{h}_1$-$\mathbf{h}_N$ plane at the ground state. The configuration of
		Bethe roots in different regions are different. Here $U=1.3$.}
	\label{fig000}
\end{figure}

In Fig.\ref{fig000}, we present the distinct regions of the boundary parameters in the $\mathbf{h}_1$-$\mathbf{h}_N$ plane.
In the different regions of boundary parameters, the patterns of Bethe roots in the reduced BAEs (\ref{BAE01}) and (\ref{BAE02}) are different, which are summarized in Table \ref{table01}.
We see that both $\{k_j\}$ and $\{\lambda_j\}$ are real or pure imaginary. The imaginary solutions correspond to the boundary strings. The boundary strings only exist in some regions of the boundary parameters.
For example, in the region ii, the boundary string is
\begin{equation}
\lambda_{M}=i\big[\frac{1-|\mathbf{h}_N|^2}{2|\mathbf{h}_N|}-\frac{U}{4}\big].
\end{equation}
Substituting $\lambda_M$ into BAEs (\ref{BAE02}), we find that
when the numerator
\begin{equation}
p+(\lambda_M-\frac{\eta}{2})|\mathbf{h}_N|\rightarrow 0,\label{BAE021}
\end{equation}
the continued product in the left hand side of BAEs (\ref{BAE02})
\begin{equation}
\prod_{l=1}^{\bar{M}}\frac{(\lambda_j+\sin k_l+i{U}/{4})(\lambda_j-\sin k_l+i{U}/{4})}{(\lambda_j+\sin k_l-i{U}/{4})(\lambda_j-\sin k_l-i{U}/{4})} \rightarrow 0,
\end{equation}
in the thermodynamic limit $\bar{M}\rightarrow \infty$.
Therefore, the boundary string $\lambda_M$ is indeed the solution of BAEs.

\begin{figure*}[t]
	\centering
	\subfigure{
		\hspace{-1.6cm}\includegraphics[scale=0.65]{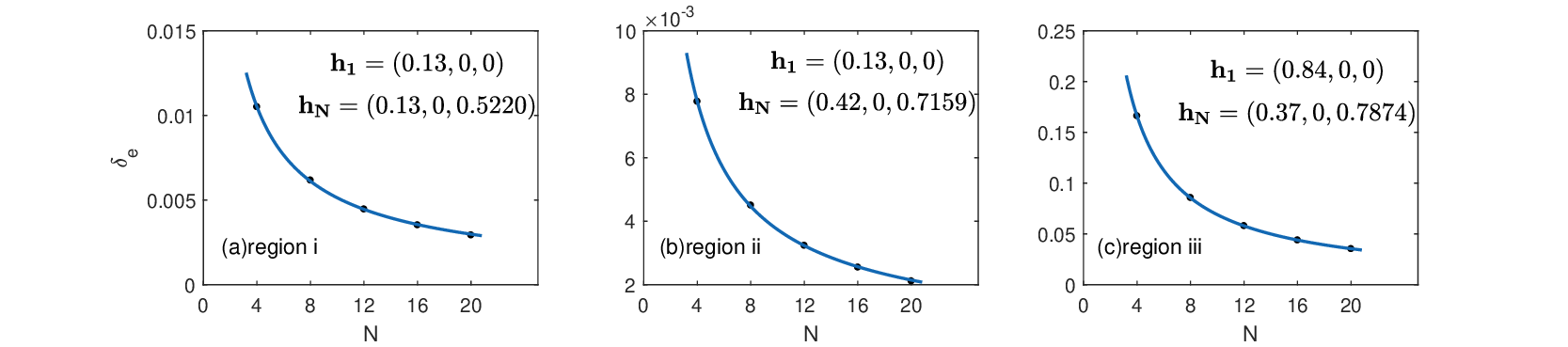}%
	}
	\hspace*{1mm}
	\subfigure{
		\includegraphics[scale=0.65]{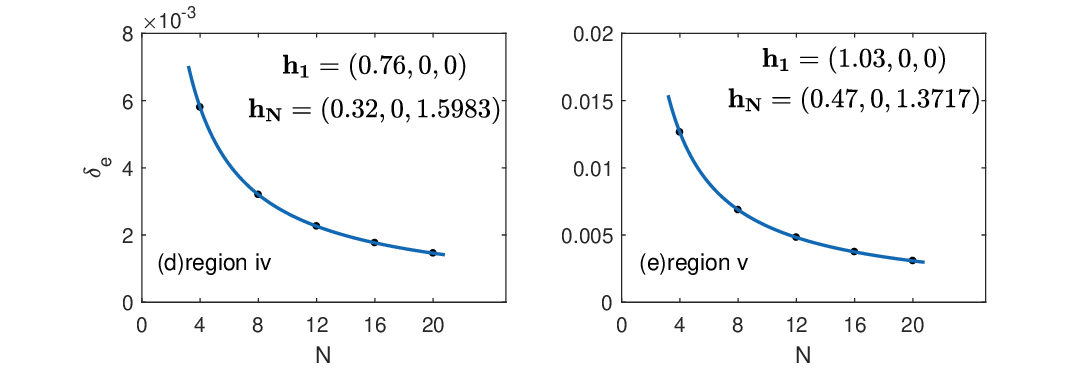}
	}
	\caption{The values of $\delta_e$ vs the system size $N$. Here the filling factor $n=\bar{M}/N=1$, the value of $E$ and $E_{hom}$ can be obtained by using the DMRG method. The value of $E$ has the general boundary parameters shown in above, the corresponding boundary parameters of $E_{hom}$ can be obtained with the help of equivalent parallel boundary fields $|\mathbf{h}_1| \hat z$ and $|\mathbf{h}_N|\hat z$. The data can be fitted as $\delta_e=\gamma N^\tau$, where (a) $\gamma$=0.0311, $\tau=-0.7822$; (b) $\gamma$=0.02358, $\tau=-0.7999$; (c) $\gamma$=0.6292, $\tau=-0.9601$; (d) $\gamma$=0.01904, $\tau=-0.8574$; (e) $\gamma$=0.04274, $\tau=-0.8783$.}
	\label{fig:inhom}
\end{figure*}

\section{Contribution of the inhomogeneous terms}
\label{sec3}

Let us now consider the contributions of the inhomogeneous term (the last term) in Eq.\eqref{BAE02-off}. For convenience, we introduce a quantity denoted as
\begin{equation}
  \delta_e=|E-E_{hom}|.\label{homo_leigenvalue}
\end{equation}
It is evident that $\delta_e$ measures the contribution of the inhomogeneous term. To determine the value of $E$ for given boundary parameters $\mathbf{h}_1=(h_1^x, h_1^y, h_1^z)$ and $\mathbf{h}_N=(h_N^x, h_N^y, h_N^z)$, we employ the DMRG method. The value of $E_{hom}$ can be obtained using the equivalent parallel boundary fields $|\mathbf{h}_1| \hat z=\sqrt{(h_1^x)^2 + (h_1^y)^2+ (h_1^z)^2}\hat z$ and $|\mathbf{h}_N|\hat z=\sqrt{(h_N^x)^2 + (h_N^y)^2+ (h_N^z)^2}\hat z$, where $\hat z$ means the unit vector along the $z$-direction.

In various regions, the distribution of Bethe roots follows different patterns, indicating that they should be considered separately. The values of $\delta_e$ as a function of the system size $N$ in different regions are depicted in Fig.\,\ref{fig:inhom}. By fitting the data, we observe that $\delta_e$ and $N$ exhibit a power-law relationship, specifically $\delta_e=\gamma N^\tau$. Notably, since $\tau<0$, we can infer that $\delta_e$ approaches zero as the system size $N$ tends to infinity.
Consequently, this implies that the physical properties of the system in the thermodynamic limit can be determined using the reduced BAEs \eqref{BAE01}-\eqref{BAE02}.
We should note that by using the fusion method and taking the XXX spin chain as an example, it has been proved that the inhomogeneous term tends to zero in the thermodynamic limit \cite{xiu-bib03}.

\section{Surface energy}
\label{sec5}

To calculate the surface energy of the model, we start by considering region i where $0<\alpha<h_0$ and $0<\beta<h_0$. In the ground state, the Bethe roots consist of $N$ real rapidities $\{k_j\}$ and $N/2$ real rapidities $\{\lambda_l\}$, with $M=N/2$. For simplicity, let's assume that $N$ is even. By substituting this distribution into the BAEs \eqref{BAE01}-\eqref{BAE02}, we obtain the following equations:
\begin{eqnarray}\label{BAE001}
  &&\frac{4(\beta\sin k_j-i\frac{\beta^2-1}{2})}
  {e^{-ik_j}-e^{ik_j}\beta^2}
  \frac{(\alpha\sin k_j-i\frac{\alpha^2-1}{2})}
  {e^{-ik_j}-e^{ik_j}\alpha^2}=e^{-2ik_jN}\nonumber\\
  &&\times\prod_{l=1}^M\frac{(\sin{k_j}+\lambda_l+i\frac{U}{4})}
  {(\sin{k_j}+\lambda_l-i\frac{U}{4})}
  \frac{(\sin{k_j}-\lambda_l+i\frac{U}{4})}
  {(\sin{k_j}-\lambda_l-i\frac{U}{4})},\\
  &&\quad j=1,\cdots,N.\nonumber
\end{eqnarray}
\begin{eqnarray}\label{BAE002}
  && \frac{\lambda_j-i\frac{U}{4}}{\lambda_j+i\frac{U}{4}}
  \frac{\lambda_j-i(\frac{1-\beta^2}{2\beta}-\frac{U}{4})}
  {\lambda_j+i(\frac{1-\beta^2}{2\beta}-\frac{U}{4})}
  \frac{\lambda_j-i(\frac{1-\alpha^2}{2\alpha}-\frac{U}{4})}
  {\lambda_j+i(\frac{1-\alpha^2}{2\alpha}-\frac{U}{4})}\nonumber\\
  &&\times\prod_{l=1}^N\frac{\lambda_j+\sin k_l-i\frac{U}{4}}
  {\lambda_j+\sin k_l+i\frac{U}{4}}
  \frac{\lambda_j-\sin k_l-i\frac{U}{4}}
  {\lambda_j-\sin k_l+i\frac{U}{4}}=\\
  &&\prod_{l=1}^M\frac{\lambda_j-\lambda_l-i\frac{U}{2}}{\lambda_j-\lambda_l+i\frac{U}{2}}
  \frac{\lambda_j+\lambda_l-i\frac{U}{2}}{\lambda_j+\lambda_l+i\frac{U}{2}},\nonumber\\
  &&j=1,\cdots,\frac{N}2.\nonumber
\end{eqnarray}
Dividing Eq.\eqref{BAE001} by its complex conjugate, we get
\begin{eqnarray}\label{BAE003}
   && \frac{(\beta\sin k_j-i\frac{\beta^2-1}{2})}
   {(\beta\sin k_j+i\frac{\beta^2-1}{2})}
   \frac{(\alpha\sin k_j-i\frac{\alpha^2-1}{2})}
   {(\alpha\sin k_j+i\frac{\alpha^2-1}{2})}\nonumber\\
   &&\times\frac{\left[\cos k_j(1-\beta^2)+i\sin k_j(1+\beta^2)\right]}
   {\left[\cos k_j(1-\beta^2)-i\sin k_j(1+\beta^2)\right]}\nonumber\\
   &&\times
   \frac{\left[\cos k_j(1-\alpha^2)+i\sin k_j(1+\alpha^2)\right]}
   {\left[\cos k_j(1-\alpha^2)-i\sin k_j(1+\alpha^2)\right]}
   =e^{-4ik_jN}\\
   &&\times \prod_{l=1}^M \left(\frac{\sin k_j+\lambda_l+i\frac{U}{4}}
   {\sin k_j+\lambda_l-i\frac{U}{4}}\right)^2
   \left(\frac{\sin k_j-\lambda_l+i\frac{U}{4}}
   {\sin k_j-\lambda_l-i\frac{U}{4}}\right)^2,\nonumber\\
   &&\quad j=1,\cdots,N.\nonumber
\end{eqnarray}

\begin{figure*}[htbp]
	\centering
	\subfigure{
		\hspace{-1.6cm}\includegraphics[scale=0.65]{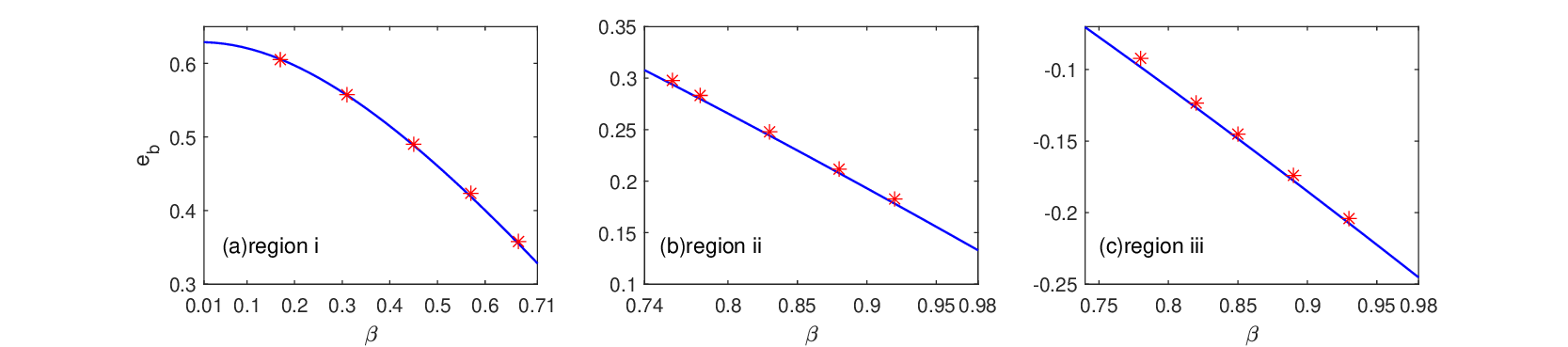}%
	}
	\hspace*{1mm} %
	\subfigure{
		\includegraphics[scale=0.65]{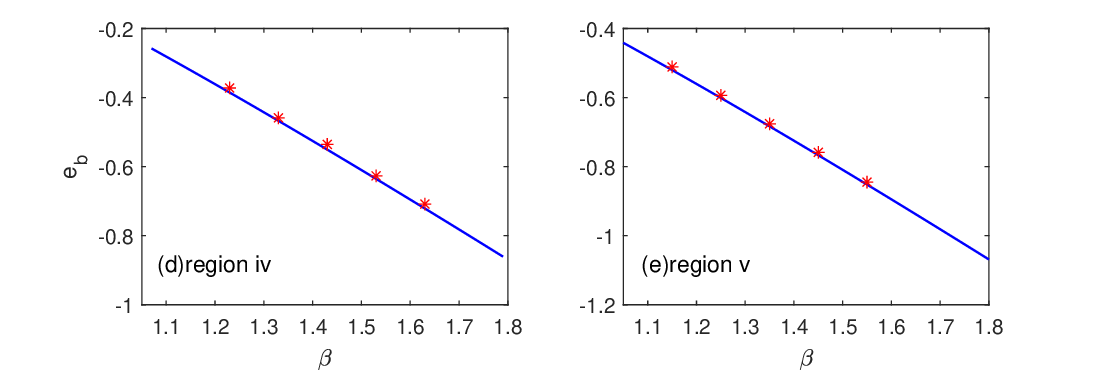}%
	}
	\caption{The surface energy $e_b$ vs the boundary $\beta$, and we
		choose $U=1.3$, so we also have $h_0=0.726$. The curve is the calculated result from Eq.\,\eqref{sur01}, and the red stars are the results obtained by using DMRG. Here we choose the red points as
(a) $\mathbf{h}_1$: (0.13,0,0), $\mathbf{h}_N$: (0.12,0,0.1204) (0.22,0,0.2184) (0.32,0,0.3164) (0.12,0,0.5572) (0.42,0,0.5220);
		(b) $\mathbf{h}_1$: (0.13,0,0), $\mathbf{h}_N$: (0.12,0,0.7505) (0.42,0,0.6573) (0.42,0,0.7733) (0.42,0,0.7159) (0.42,0,0.8185);
		(c) $\mathbf{h}_1$: (0.84,0,0), $\mathbf{h}_N$: (0.5723,0,0.53)
		(0.3315,0,0.75) (0.78,0,0.3378) (0.81,0,0.3688) (0.85,0,0.3774);
		(d) $\mathbf{h}_1$: (0.76,0,0), $\mathbf{h}_N$: (0.32,0,1.1876) (0.32,0,1.2909) (0.32,0,1.3937)
		(0.32,0,1.4962) (0.32,0,1.5983);
		(e) $\mathbf{h}_1$: (1.03,0,0),
		$\mathbf{h}_N$: (0.47,0,1.0496)
		(0.47,0,1.1583) (0.47,0,1.2655) (0.47,0,1.3717) (0.47,0,1.4770).} \label{fig:surface energy}
\end{figure*}

Taking the logarithm of Eq.\eqref{BAE003} and \eqref{BAE002}, we have
\begin{eqnarray}\label{quantum01}
  &&4\pi I_j-4\sum_{l=1}^M \arctan\frac{\sin k_j+\lambda_l}{\frac{1}{4}U}
  -4\sum_{l=1}^M \arctan\frac{\sin k_j-\lambda_l}{\frac{1}{4}U}\nonumber\\
  &&+2\arctan\frac{\cos k_j(1-\alpha^2)}{\sin k_j(1+\alpha^2)}
  +2\arctan\frac{\cos k_j(1-\beta^2)}{\sin k_j(1+\beta^2)}\\
  && -2\arctan\frac{2\alpha\sin k_j}{\alpha^2-1}
  -2\arctan\frac{2\beta\sin k_j}{\beta^2-1}=4k_jN, \nonumber\\
  &&\quad j=1,\cdots,N,\nonumber
  \end{eqnarray}
  and
  \begin{eqnarray}\label{BAE004}
   &&\sum_{l=1}^M2\arctan\frac{2(\lambda_j-\lambda_l)}{U}
   +\sum_{l=1}^M2\arctan\frac{2(\lambda_j+\lambda_l)}{U}\nonumber\\
   &&-2\arctan\frac{4\lambda_j}{U}
   -2\arctan\frac{\lambda_j}{\frac{1-\beta^2}{2\beta}
   -\frac{U}{4}}-2\arctan\frac{\lambda_j}{\frac{1-\alpha^2}{2\alpha}
   -\frac{U}{4}}\nonumber\\
   &&
   =\sum_{l=1}^N 2\arctan\frac{4(\lambda_j+\sin k_l)}{U}+
   \sum_{l=1}^N 2\arctan\frac{4(\lambda_j-\sin k_l)}{U}\nonumber\\
   &&-2\pi J_j,
   \quad j=1,\cdots,\frac{N}2.
  \end{eqnarray}
The quantum numbers, $I_j$ and $J_j$, represent the characteristics of the ground state and hold specific values
\begin{eqnarray}
\begin{aligned}
\{I_j\}&=\left\{\frac{1}{2}, \frac{3}{2},\cdots, N-\frac{3}{2}, N-\frac{1}{2}\right\}, \\
\{J_j\}&=\left\{1,2,\cdots, \frac{N}2-1, \frac{N}2\right\}.
\end{aligned}
\end{eqnarray}

In the thermodynamic limit, the distribution of Bethe roots along the real axis becomes continuous. By differentiating Eqs.\,\eqref{quantum01} and \eqref{BAE004}, we can determine the densities $\rho_c(k)$ and $\rho_s(\lambda)$ for the Bethe roots $\{k_j\}$ and $\{\lambda_l\}$ by
\begin{eqnarray}\label{rho001}
\rho_c(k)&&=\frac{1}{2\pi}+\cos k\int_{-\infty}^\infty a_1(\sin k-\lambda)\rho_s(\lambda)d\lambda+\nonumber\\
  &&\frac{1}{2N}
  [-b_\alpha(k)-b_\beta(k)+d_\alpha(k)+d_\beta(k)-\delta(k)-
\\
  &&\delta(k+\pi)/2-\delta(k-\pi)/2],\nonumber
 \end{eqnarray}
and
 \begin{eqnarray}\label{rho002}
   &&\rho_s(\lambda)+\int_{-\infty}^\infty a_2(\lambda-\lambda')d\lambda'+
   \frac{1}{2N}[ -a_1(\lambda)-a_{-\tilde{U}_\beta}(\lambda)
   -a_{-\tilde{U}_\alpha(\lambda)}]\nonumber\\
   &&\quad\quad =\int_{-\pi}^\pi a_1(\lambda-\sin k)\rho_c(k)dk
   -\\
   &&\quad\quad\quad\frac{1}{2N}[\delta(k+\infty)+\delta(k-\infty)+\delta(k)],\nonumber
\end{eqnarray}
respectively, where
\begin{eqnarray}
  a_n(x)&=&\frac{1}{\pi}\frac{4 nU}{n^2U^2+16x^2},\nonumber\\
  b_\alpha(k)&=&\frac{1}{2\pi}\frac{d}{dk}\arctan\frac{\cos k_j(1-\alpha^2)}
  {\sin k_j(1+\alpha^2)}\nonumber\\
  &=&-\frac{1}{2\pi}\frac{1}
  {\sin^2k\frac{1+\alpha^2}{1-\alpha^2}+
  \cos^2k\frac{1-\alpha^2}{1+\alpha^2}},\nonumber\\
  b_\beta(k)&=&\frac{1}{2\pi}\frac{d}{dk}\arctan\frac{\cos k_j(1-\beta^2)}
  {\sin k_j(1+\beta^2)}\nonumber\\
  &=&-\frac{1}{2\pi}\frac{1}
  {\sin^2k\frac{1+\beta^2}{1-\beta^2}+
  \cos^2k\frac{1-\beta^2}{1+\beta^2}},\nonumber\\
  d_\alpha(k)&=&\frac{1}{2\pi}\frac{2\alpha(\alpha^2-1)\cos k}
  {(\alpha^2-1)^2+(2\alpha\sin k)^2}, \nonumber\\ d_\beta(k)&=&\frac{1}{2\pi}\frac{2\beta(\beta^2-1)\cos k}
  {(\beta^2-1)^2+(2\beta\sin k)^2},\nonumber\\
  \tilde{U}_\alpha=U&+&\frac{2\alpha^2-2}{\alpha},\quad
  \tilde{U}_\beta=U+\frac{2\beta^2-2}{\beta}.\nonumber
\end{eqnarray}
The integral equations \eqref{rho001}-\eqref{rho002} can be solved via the Fourier transformation. From Eq.\eqref{rho002}, we obtain
\begin{eqnarray}\label{four001}
  &&(1+\tilde{a}_2(\omega))\tilde{\rho}_s(\omega)=
  \int_{-\pi}^\pi e^{i\omega\sin k}e^{-\frac{U}{4}|\omega|}\rho_c(k)dk+\nonumber\\
  &&\frac{1}{2N}\left(e^{-\frac{U|\omega|}{4}}+e^{\frac{\tilde{U}_\beta|\omega|}{4}}
+e^{\frac{\tilde{U}_\alpha|\omega|}{4}}-1-e^{i\omega\infty}-e^{-i\omega\infty}\right),
\end{eqnarray}
where $\tilde{a}_n(\omega)=e^{-nU|\omega|/4}$. Substituting Eq.\eqref{rho001} into the above equation, we have
\begin{eqnarray}
  &&\tilde{\rho}_s(\omega)=\int_{-\pi}^\pi\frac{e^{i\omega\sin k}}{2\cosh\frac{U}{4}|\omega|}\left\{
  \frac{1}{2\pi}+\frac{1}{2N}\left[-b_\alpha(k)-b_\beta(k)+\right.\right.
  \nonumber\\
  &&\left.\left.d_\alpha(k)+
  d_\beta(k)-\delta(k)-\frac{1}{2}\delta(k+\pi)-\frac{1}{2}\delta(k-\pi)
  \right]\right\}dk+\\
  &&\frac{1}{2N}
  \frac{e^{-{U}|\omega|/{4}}+e^{{\tilde{U}_\beta}|\omega|/{4}}
  +e^{{\tilde{U}_\alpha}|\omega|/{4}}-1-e^{i\omega \infty}
  -e^{-i\omega\infty}}{1+e^{-{U}|\omega|/{2}}},\nonumber
\end{eqnarray}
which gives
\begin{eqnarray}\label{rho 002-2}
  &&\rho_s(\mu) = \frac{1}{2\pi}\int_{-\infty}^\infty
  \frac{e^{-i\omega\mu}}{2\cosh\frac{U}{4}|\omega|}J_0(\omega)d\omega
  +\nonumber\\
  &&\quad\quad\frac{1}{2N}\frac{1}{2\pi}\int_{-\infty}^\infty
  \frac{e^{-i\omega\mu}}{2\cosh\frac{U}{4}|\omega|}\int_{-\pi}^\pi
  e^{i\omega\sin k}\left[-b_\alpha(k)-b_\beta(k)-\right.\nonumber\\
  &&\quad\quad\left.\delta(k)-\frac{1}{2}\delta(k-\pi)
   -\frac{1}{2}\delta(k+\pi)
  \right]dk d\omega+\\
   &&\quad\quad\frac{1}{2N}\frac{1}{2\pi}\int_{-\infty}^\infty
   e^{-i\omega\mu}\frac{e^{-\frac{U}{4}|\omega|}+e^{\frac{\tilde{U}_\beta}{4}|\omega|}
  +e^{{\tilde{U}_\alpha}|\omega|/{4}}}{1+e^{-{U}|\omega|/{2}}}d\omega+\nonumber\\
  &&\quad\quad\frac{1}{2N}\frac{1}{2\pi}\int_{-\infty}^\infty
   e^{-i\omega\mu}\frac{-1-e^{i\omega \infty}
  -e^{-i\omega\infty}}{1+e^{-{U}|\omega|/{2}}}d\omega,\nonumber
\end{eqnarray}
where $J_0(\omega)=\frac{1}{2\pi}\int_{-\pi}^\pi e^{i\omega \sin k}dk$. Substituting Eq.\eqref{rho 002-2} into \eqref{rho001}, we have
\begin{eqnarray}
  \rho_c(k) &=& \frac{1}{2\pi}+\frac{\cos k}{2\pi}\int_{-\infty}^\infty
  \frac{e^{i\omega\sin k}e^{-\frac{U}{4}|\omega|}}{2\cosh\frac{U}{4}|\omega|}
  J_0(\omega)d\omega
  \nonumber\\
  &+&
  \frac{1}{2N}\left[-b_\alpha(k)-b_\beta(k)+d_\alpha(k)+d_\beta(k)\right]
  \nonumber\\
  &+&
  \frac{1}{2N}\left[
  -\delta(k)-\frac{\delta(k+\pi)}{2}-\frac{\delta(k-\pi)}{2}\right]
  \\
  &+&\frac{1}{2N}\frac{\cos k}{2\pi}\int
  _{-\infty}^\infty
  \frac{e^{i\omega\sin k}e^{-\frac{U}{4}|\omega|}}{2\cosh\frac{U}{4}|\omega|}
  \left(-\tilde{b}_\alpha(\omega)-\tilde{b}_\beta(\omega)\right.\nonumber\\
  &+&\left.e^{-\frac{U}{4}|\omega|}+e^{\frac{\tilde{U}_\beta}{4}|\omega|}
  +e^{\frac{\tilde{U}_\alpha}{4}|\omega|}
  -3-e^{i\omega\infty}-e^{-i\omega\infty}\right)d\omega,\nonumber
\end{eqnarray}
where
\begin{eqnarray}
  \int_{-\pi}^\pi e^{i\omega\sin k}b_\alpha(k)dk=\tilde{b}_\alpha(\omega),\quad\int_{-\pi}^\pi e^{i\omega\sin k}b_\beta(k)dk=\tilde{b}_\beta(\omega).\nonumber
\end{eqnarray}
The ground state energy of the system thus reads
\begin{eqnarray}
\begin{aligned}
  E_g &= -2N\int_{-\pi}^{\pi}\cos k \rho_c(k)dk\\
  &=-4N \int_0^\infty
  \frac{J_0(\omega)J_1(\omega)}{\omega\left(1+e^{\frac{U\omega}{2}}\right)}
  d\omega+e_b,
\end{aligned} \label{ww}
\end{eqnarray}
where
\begin{eqnarray}
  J_1(\omega)=\frac{\omega}{2\pi}\int_{-\pi}^\pi
  \cos k^2\cos (\omega \sin k)dk,\nonumber
\end{eqnarray}
and $e_b$ is the energy induced by the boundary magnetic fields. The definition of surface energy is \cite{xiu-bib04}
\begin{equation}\label{1}
e_b=\lim_{N \rightarrow \infty} \left(E_g- N e_{\infty}\right),
\end{equation}
where $E_g$ is the ground state energy of the system with open boundary condition, and $e_{\infty}$ is the ground state energy density of the corresponding periodic system.
Because the terms with the order $O(N)$ in Eq.(\ref{1}) are canceled with each other, the result is a boundary parameters dependent constant.
Thus, the surface energy is a well-defined physical quantity that is worth studying. From Eqs.(\ref{ww}) and (\ref{1}), we obtain the surface energy
\begin{eqnarray}
  e_b&=&-\int_{-\pi}^{\pi}\frac{\cos k^2}{2\pi}
  \int_{-\infty}^\infty \frac{e^{i\omega\sin k}}{1+e^{\frac{U\omega}{2}}}
  \left(-\tilde{b}_\alpha(\omega)-\tilde{b}_\beta(\omega)\right)d\omega dk
  \nonumber\\
  &&+2\int_{-\pi}^{\pi}\frac{\cos k^2}{2\pi}
  \int_{-\infty}^\infty \frac{e^{i\omega\sin k}}{1+e^{\frac{U\omega}{2}}}
   d\omega dk
  \nonumber\\
  &&-\int_{-\pi}^\pi \frac{\cos k^2}{2\pi}\int
  _{-\infty}^\infty\frac{e^{i\omega\sin k}}{2\cosh \frac{U}{4}|\omega|}
  \left[e^{-\frac{U}{4}|\omega|}-1\right.\nonumber\\
  &&\left.-e^{i\omega\infty}-e^{-i\omega\infty}+f_1(\omega,\alpha,\beta)
  \right]d\omega dk\nonumber\\
  &&-\int_{-\pi}^\pi\cos k\left[d_\alpha(k)+d_\beta(k)\right] dk\nonumber\\
  &&-\int_{-\pi}^\pi\cos k f_2(k,\alpha,\beta)dk,
  \label{sur01}
\end{eqnarray}
with
\begin{eqnarray}
  f_1(\omega,\alpha,\beta)=
  e^{\frac{\tilde{U}_\beta}{4}|\omega|}
  +e^{\frac{\tilde{U}_\alpha}{4}|\omega|} \quad f_2(k,\alpha,\beta)=0.\nonumber
\end{eqnarray}

Using the same technique, we can calculate the surface energy in other regions. These regions exhibit the same form as Eq\,\eqref{sur01}, with differences arising from the expressions of $f_1(\omega, \alpha, \beta)$ and $f_2(k, \alpha, \beta)$.
\begin{numcases}{f_1(\omega,\alpha,\beta)=}
-e^{\frac{-\tilde{U}_\beta}{4}|\omega|}
  +e^{\frac{\tilde{U}_\alpha}{4}|\omega|} \nonumber\\
  -e^{-\tilde{U}_{01}^\beta|\omega|}-e^{\tilde{U}_{02}^\beta|\omega|},
  &  $ii$ \nonumber\\
  -e^{\frac{-\tilde{U}_\beta}{4}|\omega|}
  -e^{\frac{-\tilde{U}_\alpha}{4}|\omega|}
  -e^{-\tilde{U}^\alpha_{01}|\omega|}\nonumber\\
  -e^{\tilde{U}^\alpha_{02}|\omega|}
  -e^{-\tilde{U}^\beta_{01}|\omega|}-e^{\tilde{U}^\beta_{02}|\omega|}, &  $iii$\nonumber\\
  e^{\frac{\tilde{U}_\alpha}{4}|\omega|}
  -e^{\tilde{U}_{02}^\beta|\omega|}, &  $iv$\nonumber\\
  -e^{\tilde{U}_{02}^\alpha|\omega|}
  -e^{\tilde{U}_{02}^\beta|\omega|}, &  $v$\nonumber
\end{numcases}
\begin{numcases}{f_2(k,\alpha,\beta)=}
  b_{01}(k)-b_{02}(k), &  $ii$ \nonumber\\
b^\alpha_{01}(k)-b^\alpha_{02}(k)+b^\beta_{01}(k)-b^\beta_{02}(k), &  $iii$\nonumber\\
b_{01}(k)-b_{02}(k), &  $iv$\nonumber\\
b_{01}^\alpha(k)-b_{02}^\alpha(k)
  +b_{01}^\beta(k)-b_{02}^\beta(k), &  $v$\nonumber
\end{numcases}
where
\begin{eqnarray}
  && b_{01}(k)=\frac{1}{\pi}\frac{\left(\lambda_0+\frac{U}{4}\right)\cos k}
  {\left(\lambda_0+\frac{U}{4}\right)^2+\sin ^2k}, \nonumber\\
  && b_{02}(k)=\frac{1}{\pi}\frac{\left(\lambda_0-\frac{U}{4}\right)\cos k}
  {\left(\lambda_0-\frac{U}{4}\right)^2+\sin ^2k},\nonumber\\
  &&b_{01}^\alpha(k)=\frac{1}{\pi}\frac{\left(\lambda_{01}+\frac{U}{4}\right)\cos k}
  {\left(\lambda_{01}+\frac{U}{4}\right)^2+\sin ^2k},\nonumber\\
  &&b_{02}^\alpha(k)=\frac{1}{\pi}\frac{\left(\lambda_{01}-\frac{U}{4}\right)\cos k}
  {\left(\lambda_{01}-\frac{U}{4}\right)^2+\sin ^2k},\nonumber\\
  &&
   b_{01}^\beta(k)=\frac{1}{\pi}\frac{\left(\lambda_{02}+\frac{U}{4}\right)\cos k}
  {\left(\lambda_{02}+\frac{U}{4}\right)^2+\sin ^2k}, \nonumber\\
  &&b_{02}^\beta(k)=\frac{1}{\pi}\frac{\left(\lambda_{02}-\frac{U}{4}\right)\cos k}
  {\left(\lambda_{02}-\frac{U}{4}\right)^2+\sin ^2k},\nonumber\\
  && \lambda_0=\frac{1-\beta^2}{2\beta}-\frac{U}{4}, \nonumber\\
  && \lambda_{01}=\frac{\alpha^2-1}{2\alpha}+\frac{U}{4}
 \quad \lambda_{02}=\frac{\beta^2-1}{2\beta}+\frac{U}{4},\nonumber\\
  &&\tilde{U}^\beta_{01} = U+\frac{2-2\beta^2}{\beta} \quad
  \tilde{U}^\beta_{02} = -3U+\frac{2-2\beta^2}{\beta},\nonumber\\
    &&\tilde{U}^\alpha_{01} = U+\frac{2-2\alpha^2}{\alpha} \quad
  \tilde{U}^\alpha_{02} = -3U+\frac{2-2\alpha^2}{\alpha}.\nonumber
\end{eqnarray}

To verify the accuracy of the analytical expression \eqref{sur01}, we also compute the surface energy using the DMRG approach. The results, depicted as red stars in Fig\,\ref{fig:surface energy}, are compared to the surface energies obtained from the analytical expression \eqref{sur01}, represented by the blue curves in the same figure. It is evident that there is excellent agreement between the analytical and DMRG results. However, it is important to note that the surface energy becomes infinite if either of the boundary parameters $\mathbf{h}_1$ or $\mathbf{h}_N$ tends to infinity, as in this case the Hamiltonian \eqref{Ham} is dominated by the boundary fields.

From the calculation in this paper, we see that the energy of the system with parallel and that with nonparallel boundary magnetic fields are the same up to the order $O(N^{0})$.
Therefore, if we study the surface energy and other physical quantities with the order $O(N^{0})$, the parallel and nonparallel boundary magnetic fields have no difference. This means that in the thermodynamic limit,
the contributions of two boundary fields are independent. The ground state energy with nonparallel boundary fields can be obtained by using the equivalent parallel ones.
We note that all the components in the boundary fields are included in the expression of energy.

However, the energies with the order $O(N^{-1})$ for the parallel and nonparallel boundary magnetic fields would be different, which can be calculated from the integral equations (17)-(18) with the order $O(N^{-2})$.

We also note that the eigen-states of the system with nonparallel boundary magnetic fields are totally different from that with equivalent parallel ones in the thermodynamic limit. The boundary magnetic fields can affect the spin configurations of electrons in the bulk \cite{xiu-bib022}.
The eigenstates of the system are helical, which are quite different from those with parallel boundary fields or periodic boundary condition. By using the generalized algebraic Bethe ansatz or the separation of variables,
the eigenstates of the system can be written out explicitly.

\section{Conclusion}\label{sec6}
In this study, we investigate the surface energy of the one-dimensional Hubbard model with nonparallel boundary magnetic fields. At zero temperature, the distribution of Bethe roots can be classified into five regions based on varying boundary parameters, indicating that the nonparallel boundary fields lead to five distinct ground states. By analyzing the patterns of Bethe roots, the ground state energy and surface energy of the system can be computed. The findings demonstrate the existence of stable boundary strings under specific boundary fields. These results provide a sound foundation for calculating the spin expectation value and spin-spin correlation function considering different regions of the boundary magnetic fields, as well as for generating excitations in the bulk through the addition of spinons, strings, quartets, and other entities. Notably, further exploration of these results at finite temperatures, along with investigating the finite size effect and patterns of the inhomogeneous BAEs, will serve as interesting future research topics.

\acknowledgments
We thank Professor Y. Wang for his valuable discussions and continuous encouragement. P.S thanks Dr. Kang Wang for valuable discussions. We acknowledge the financial support from the National Key R$\&$D Program of China (Grant No.2021YFA1402104), National Natural Science Foundation of China (Grant Nos. 12074410, 12175180, 12205235, 12247103, 12247179 and 11934015), the Major Basic Research Program of Natural Science of Shaanxi Province(Grant Nos. 2021JCW-19 and 2017ZDJC-32), China Postdoctoral Science Foundation (Grant No. 2022M712580), Scientific Research Program Funded by Education Department of Shaanxi Provincial Government (Grant Nos. 22JK0577), Shaanxi Fundamental Science Research Project for Mathematics and Physics (Grant Nos. 22JSZ005), and the Strategic Priority Research Program of the Chinese Academy of Sciences (Grant No. XDB33000000).

\end{document}